\documentclass[prl,preprint, showpacs,amssymb,superscriptaddress]{revtex4-1}

\usepackage{graphicx, color}

\newcommand{\Let}{ \mathrel{\mathop : \! \! = } }

\begin{document}
\title{Experimental detection of non-classical correlations in mixed state quantum computation}

\author{G. Passante}
\affiliation{Institute for Quantum Computing and Dept. of Physics, University of Waterloo, Waterloo, ON, N2L 3G1, Canada.}

\author{O. Moussa}
\affiliation{Institute for Quantum Computing and Dept. of Physics, University of Waterloo, Waterloo, ON, N2L 3G1, Canada.}

\author{D.A. Trottier }
\affiliation{Institute for Quantum Computing and Dept. of Physics, University of Waterloo, Waterloo, ON, N2L 3G1, Canada.}

\author{R. Laflamme}
\affiliation{Institute for Quantum Computing and Dept. of Physics, University of Waterloo, Waterloo, ON, N2L 3G1, Canada.}
\affiliation{Perimeter Institute for Theoretical Physics, Waterloo, ON, N2J 2W9, Canada}

\date{\today}

\begin{abstract}
We report on an experiment to detect non-classical correlations in a highly mixed state.  The correlations are characterized by the quantum discord and are observed using four qubits in a liquid state nuclear magnetic resonance quantum information processor.  The state analyzed is the output of a DQC1 computation, whose input is a single quantum bit accompanied by $n$ maximally mixed qubits.  This model of computation outperforms the best known classical algorithms, and although it contains vanishing entanglement it is known to have quantum correlations characterized by the quantum discord.  This experiment detects non-vanishing quantum discord, ensuring the existence of non-classical correlations as measured by the quantum discord. 
\end{abstract}


\maketitle

Entanglement has often been synonymous with {\it quantum}.  Over the past several decades, entanglement has been well studied in the context of quantum information and computation as it plays a crucial role in many protocols, such as teleportation and superdense coding~\cite{Nielsen2000Quantum-Computa}. It has been shown as a vital component to the speedup exhibited by pure state quantum computation~\cite{Jozsa1996On-the-role-of-}. However, in the context of mixed state quantum computation it is not known what role entanglement plays.  In fact, there is a computational model that contains limited entanglement~\cite{Datta2005Entanglement-an}, yet an advantage over current classical methods appears to exist~\cite{Knill1998Power-of-One-Bi}.  This model of mixed state quantum computation is known as DQC1, or deterministic quantum computation with one quantum bit, and has been of great interest in recent years. 

While DQC1 contains very little, or no entanglement, it does contain non-classical correlations as measured by the quantum discord~\cite{Datta2007Quantum-discord}.  Quantum discord is a measure of the correlations that exist in excess of those present in classical states.  It is measured by the difference in two classically equivalent formulations of the mutual information, where a non-zero value indicates a deviation from purely classical correlations~\cite{Henderson2001Classical-quant,Ollivier2002Quantum-Discord}.  It is not yet known whether or not quantum discord assists quantum algorithms, but it is a good candidate for the computational advantage offered by the DQC1 model, and understanding it better will undoubtably provide insights into the workings of quantum systems and algorithms.  While it has been shown that almost every quantum state has quantum discord~\cite{Ferraro:2010fk}, Datta et al. showed that on average, the quantum discord present in a DQC1 algorithm drops with a decrease in polarization~\cite{Datta2007Quantum-discord}. Typical liquid state nuclear magnetic resonance (NMR) experiments are performed at room temperature, and the unentangled initial states have very small polarization.  Therefore, the question remains, is it possible to experimentally detect quantum discord in a DQC1 algorithm where the polarization is very small?

We report on the experimental detection of non-classical correlations in the output state of a DQC1 algorithm in NMR using a state-independent non-zero discord witness.  Similar to entanglement, calculation of the quantum discord requires full tomographic data, therefore witnesses that detect the existence of these correlations are a practical alternative.  Before describing our results in detail, we define both the DQC1 model and quantum discord. We then explain the method used to detect non-classical correlations, followed by a description of the physical system.  Finally, we conclude by discussing the significance of these results.

\textbf{DQC1}: Deterministic quantum computation with one quantum bit (often misnamed {\it one clean qubit}) is a model of computation where only one qubit is slightly polarized away from the maximally mixed state.  
\begin{figure}[h]
\includegraphics[scale=.35]{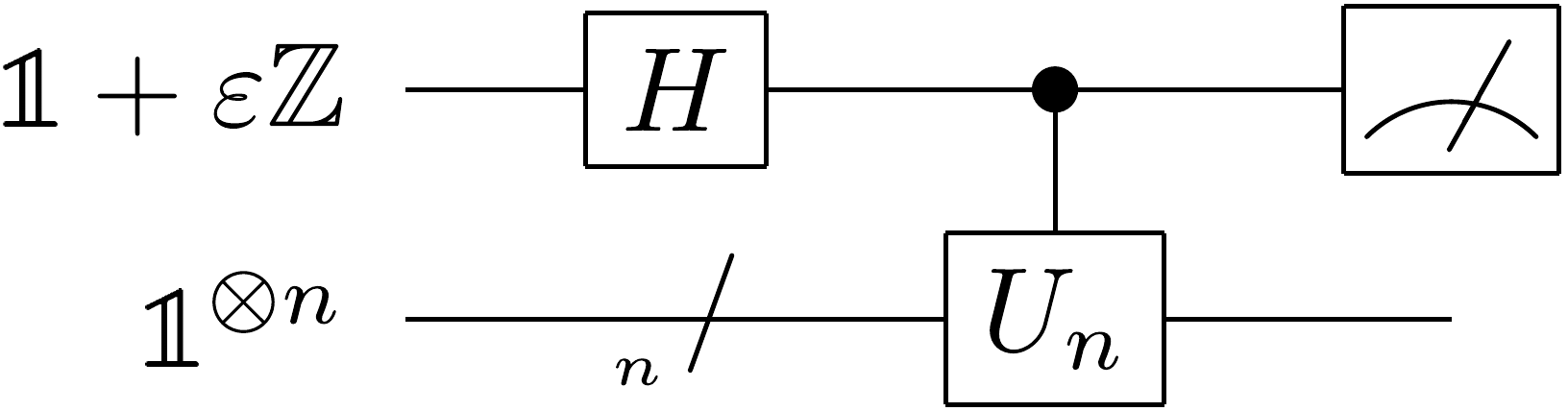}
\caption{The DQC1 circuit where the top qubit has a bias of $\varepsilon$ towards the ground state.  Measurements of $\langle \sigma_x \rangle$ and $\langle \sigma_y \rangle$ yield the real and imaginary parts of $\varepsilon \mbox{Tr}(U_n)/2^n$, respectively.  In the initial state, the identity terms are recognized as properly normalized states such that $\mbox{tr}(\rho) = 1$.}
\label{dqc1}
\end{figure}
The DQC1 circuit seen in Figure \ref{dqc1} calculates the real and imaginary components of the trace of the unitary $U_n$ upon measurement of the expectation values of Pauli operators $\sigma_x$ and $\sigma_y$, respectively.  Estimating the trace of a unitary matrix is a problem with no known efficient classical algorithm, thus, a DQC1 quantum computer outperforms its current classical counterparts.  Interestingly, there is never entanglement between the top qubit and the rest of the system.  With any other bipartite split there can be a small amount of entanglement that does not grow with the size of the system~\cite{Datta2005Entanglement-an}, suggesting that entanglement may not be the cause of the quantum speedup in DQC1.  Despite the small amount of entanglement found in this model, it contains a significant amount of non-classical correlations as measured by quantum discord for a randomly chosen unitary~\cite{Datta2007Quantum-discord}.  

\textbf{Discord}:  Quantum discord measures non-classical correlations in bipartite states that are not necessarily captured by entanglement~\cite{Henderson2001Classical-quant,Ollivier2002Quantum-Discord}. 
These non-classical correlations include, but are not limited to, entanglement, and are measured by the difference of two expressions for the mutual information, $I(A \mathop : B) \Let H(A) + H(B) - H(A,B)$ and $J(A \mathop : B) \Let H(B) - H(B|A)$. For states described by a classical probability $\{p_i\}$,  $H(A) = -\sum_i p_i \log{p_i}$ is the Shannon entropy, and $H(B|A)$ is the conditional entropy, averaging the entropy of system $B$ over the different outcomes on system $A$~\cite{Cover:2006fk}.  If $A$ and $B$ refer to quantum mechanical systems, then their states are described by operators on Hilbert spaces  $\cal{H}_{AB} = \cal{H}_A \otimes \cal{H}_B$ and $H(A) = -\mbox{Tr}(\rho_A \log{\rho_A})$ is the von Neumann entropy.  Classically the two expressions for the mutual information are equivalent, $I(A \mathop : B) = J(A\mathop : B)$. However, if all possible measurements on a subsystem of the bipartite state disturb the quantum state in $\cal{H}_{AB}$, then there exist correlations that are not found in completely classical systems, and $I(A\mathop : B) \neq J(A\mathop : B)$.
The quantum discord $D(A\mathop : B)$ is defined as the minimum difference between the two formulations of the mutual information, and reduces to
\begin{equation}
D(A\mathop : B)= H(\rho_A) - H(\rho)  + \mbox{min}_{\{E_k\}} \sum_k p_k H(\rho_{B|k})\,,
\label{discord}
\end{equation} 
where $\rho_A = \mbox{Tr}_B(\rho)$ is the reduced density matrix of system $A$, $\{E_k\}$ is a complete set of orthonormal projectors on $\cal{H}_A$ such that $\sum_k E_k = \mathbb{I}$, $p_k$ is the probability of observing outcome $k$ on system $A$, and $\rho_{B|k} \Let \mbox{Tr}_A((E_k \otimes \mathbb{I}_B) \rho)$ is the state of system $B$ conditional on the measurement of system $A$ returning measurement outcome $k$.

Employing Equation~\ref{discord}, the amount of discord in a given state can be estimated relying on a full description of the bipartite state.  In a recent experiment, full state tomography from a two qubit optical implementation of DQC1 was used to estimate the quantum discord~\cite{Lanyon2008Experimental-qu}.  On the other hand, to {\it detect} non-zero quantum discord, there exist methods that do not require full state tomography.  The experiment presented in this letter follows one such proposal of a state independent non-zero discord witness~\cite{dakic2010necessary}.  The procedure shows that by writing the state of the bipartite system as $\rho_{AB} = \sum_{n,m} r_{nm} A_n \otimes B_m$, where $\{A_n\}$ and $\{B_m\}$ are bases of Hermitian operators and $r_{nm}$ are the matrix elements of the {\it correlation matrix} $R$, the quantum discord $D(A\mathop : B)$ is non-zero if the rank of the correlation matrix $R$ is greater than the dimension of system A (similarly, $D(B\mathop : A)$ is non-zero if $rank(R)>dim(B)$).  That is to say that in order to show that a system has non-zero discord the number of measurements needed is only as many as is needed to show that $rank(R) > dim(A)$.

In our experiment, we find the rank by measuring several columns of $R$ and examining the singular values to determine a lower bound on the rank.  If this lower bound is greater than the dimension of system $A$, we can conclude that the system contains non-zero discord.  If the lower bound is less than or equal to the dimension of system $A$, then we measure an additional column and find the singular values again.  We continue this procedure until either the rank of $R$ is greater than the dimension of $A$, or full state tomography has been performed, in which case this procedure produces no conclusive statement regarding the presence or absence of quantum discord.  

\textbf{Experimental System}: We implement this algorithm using a nuclear magnetic resonance quantum information processor in the liquid state.  NMR computation  at room temperature is highly mixed and thus, is an ideal system for implementing the DQC1 model of computation.  The state of the system in NMR is written as $\rho_{N\!M\!R} = (1-\alpha)I / 2^N+ \alpha\rho_{pps}$, where $\rho_{pps}$ is known as the pseudopure state with unit trace and $N$ is the total number of qubits.  Note that $\rho_{pps}$ need not be pure and the name is used for historical reasons.  Measurements of $\rho_{pps}$ are compared to a known state that serves as a reference for $\alpha$.  At thermal equilibrium, $\alpha$ is the ground state bias as given by the Boltzmann distribution and is equal to $\hbar \gamma B^0 / 2k_B T$, where $B^0$ is the value of the static magnetic field, $\gamma$ is the gyromagnetic ratio, and $k_BT$ is the thermal energy. For our experiment, with carbon-13 nuclei at room temperature in a 16.4 T magnet, the polarization is $1.4 \times 10^{-5}$.  For this polarization and perfect implementation of the unitary transformations, the numerically computed discord present at the completion of the algorithm is $5.4 \times 10^{-11}$. Analytical results~\cite{Datta2007Quantum-discord} for the average discord after a DQC1 circuit for a unitary drawn uniformly by the Haar measure indicate a discord of approximately $7.1 \times 10^{-11}$.

The existence of non-classical correlations in an NMR state, as measured by the quantum discord, does not depend on the value of the polarization, provided it is non-zero -- this can be deduced as follows: It has been shown~\cite{A.Datta:2008kx} that a bipartite state has zero discord $D(A\mathop : B)$, if and only if there exists a projective measurement $\{E_k\}$ on system $A$ such that 
\begin{equation}
\rho_{AB} = \sum_k (E_k \otimes \mathbb{I}) \rho_{AB} (E_k \otimes \mathbb{I}).
\label{zero discord}
\end{equation}  
Noting that equation (\ref{zero discord}) is true for $\rho_{N\!M\!R}$ if and only if it is true for $\rho_{pps}$, indicates that if non-zero discord is detected for $\rho_{pps}$ then the full NMR state also contains non-zero discord.

\begin{figure}[h]
\includegraphics[scale=.097]{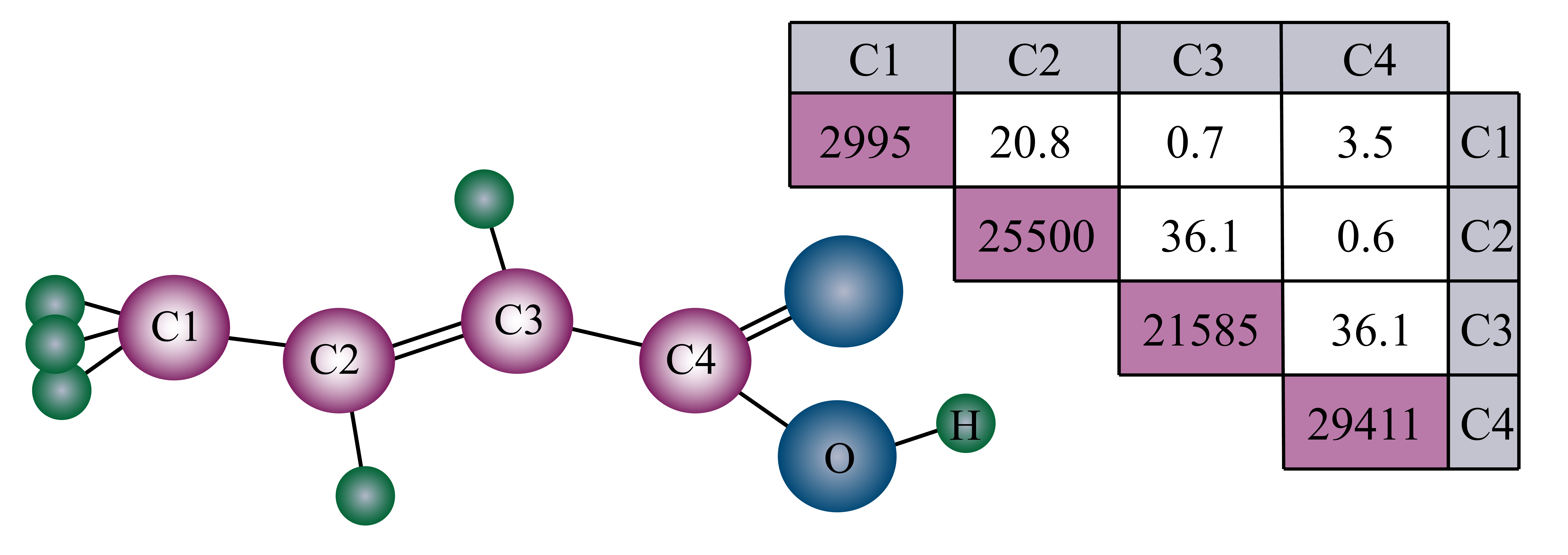}
\caption{(Color online) The molecule trans-crotonic acid and a table with the parameters of the Hamiltonian given in Hz.  The shaded diagonal elements represent the chemical shifts $\omega_i$ with the Hamiltonian $\Sigma_i \pi\omega_i\sigma_z^i$.  The remaining elements indicate the scalar coupling constants $J_{ij}$ with the Hamiltonian $\Sigma_{i<j}\pi J_{ij}\sigma_z^i\sigma_z^j$.}
\label{molecule}
\end{figure}

We experimentally test for the existence of quantum discord in a liquid state NMR implementation of a four qubit DQC1 algorithm.  
Our qubits are the four carbon nuclei in the carbon-13 labelled molecule trans-crotonic acid (see Figure \ref{molecule}).  The qubits are a bulk ensemble of approximately $10^{20}$ identical spin-1/2 nuclei that are manipulated in parallel and exhibit a two level energy splitting in the presence of high magnetic fields.  Measurements of $\langle\sigma_x\rangle$ and $\langle\sigma_y\rangle$ are performed using quadrature detection of the free induction decay.

The experiment is implemented in a Bruker Avance 700 MHz spectrometer where the hydrogen nuclei are always decoupled using the WALTZ-16 and bi-WALTZ-16 composite pulse sequences \cite{Shaka1983Evaluation-of-a}.  Radio frequency (r.f.)~pulses in the plane perpendicular to the static magnetic field are numerically generated using the GRAPE algorithm \cite{Khaneja2005Optimal-control,Ryan:2008fk}, which starts from a random initial guess and is iteratively improved through a gradient ascent search.  These pulses are optimized to start and finish with zero amplitude, produce a fidelity $|\text{tr}(U^\dagger_{goal} U_{sim})|^2/d^2$, where $d$ is the dimension of the Hilbert space of $U_{goal}$, of at least 0.998, and be robust against r.f.~inhomogeneities ($\pm 3\%$).  Before running the experiment, the pulses are adjusted for non-linearities in pulse generation and transmission by placing a pickup coil inside the bore of the magnet and running a feedback loop, iteratively making small changes to the pulse shape to ensure as close to perfect transmission as possible. After the feedback loop, pulses are found to have a simulated fidelity of approximately $0.99$ with respect to the ideal unitary.

In this experiment, the bipartite split is the natural choice of the first, slightly polarized qubit, and the remaining three.  Thus, our state can be represented as $\rho_{N\!M\!R} =(1-\alpha)I^{\otimes 4}/2^4 + \alpha( \sum_{n,m} r_{nm} A_n \otimes B_m)$, where only $\rho_{pps}$ has been written in terms of Hermitian observables.  We choose the observables to be $\{A_n\}=\{ I, X, Y, Z\}$ and $\{B_m\} = \{ III, IIX, IIY, IIZ, IXX, \ldots\}$, where $X = \sigma_x$ (similarly for $Y$ and $Z$) and $\{B_m\}$ are all possible combinations of three qubit Pauli operators.  The size of our correlation matrix $R$ is then $4\times64$ with a maximum rank of 4, and non-zero quantum discord $D(A \mathop : B)$ is witnessed when $rank(R) > dim(A) = 2$.  The unitary transformation used in this instance of the DQC1 model is $U = \mbox{diag}(a,a,b,1,a,b,1,1)$, where $a = -(e^{-i3\pi/5})^4$ and $b = (e^{-i3\pi/5})^8$, and is an important transformation in the approximation of the Jones polynomial for a class of knots whose braid representations have four strands.  The problem of evaluating the Jones polynomial is not efficient using current classical methods, but has recently been shown to completely encapsulate the power of the DQC1 model \cite{Shor2007Estimating-Jone} and has been experimentally implemented to successfully distinguish a specific class of knots \cite{passante2009experimental}.

\begin{figure}[h]
\includegraphics[scale=0.35]{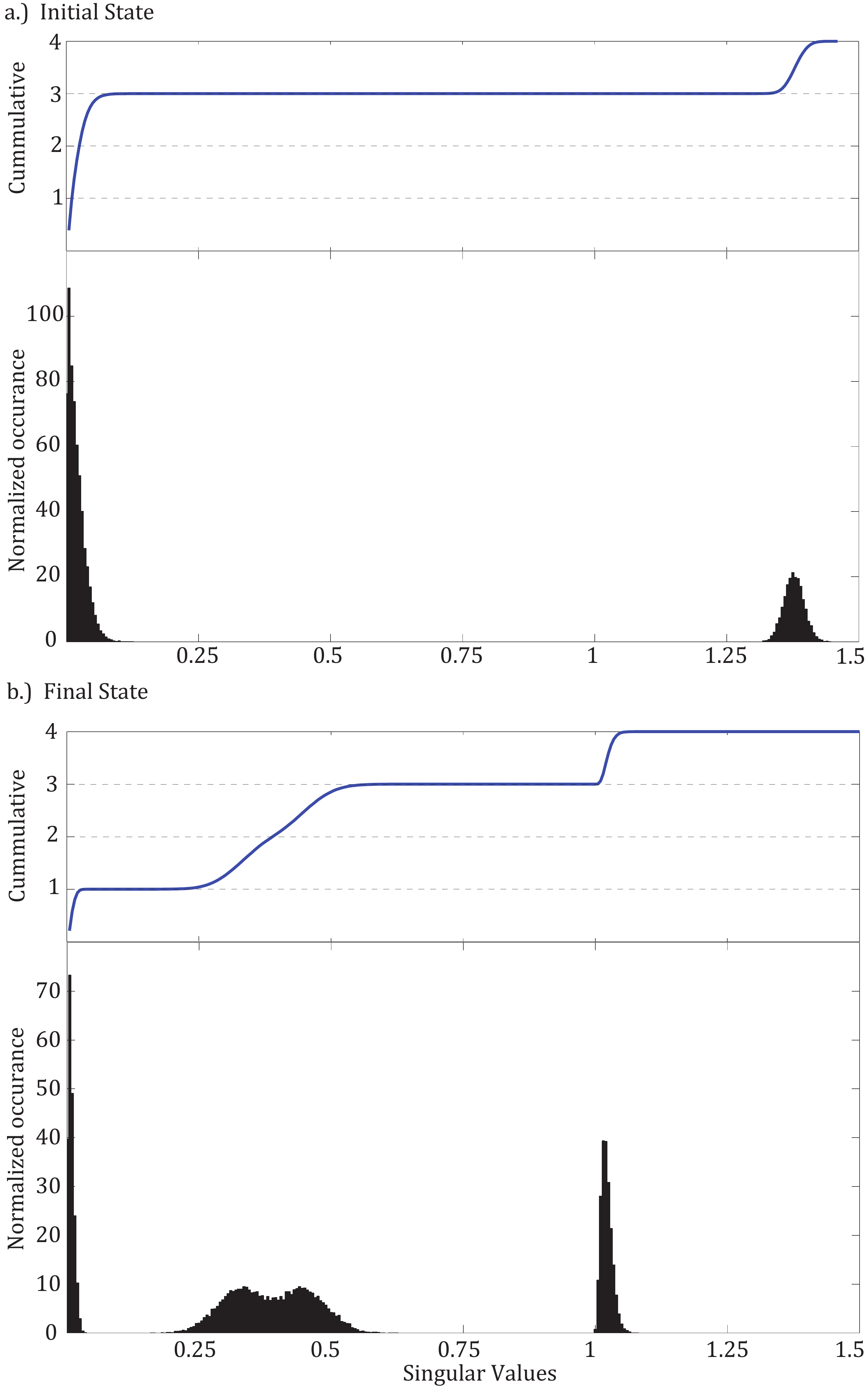}
\caption{Shown are the distributions of the singular values (and their cumulative distributions) computed for the experimentally determined correlation matrices of the (a) initial and (b) final states of a DQC1 algorithm implemented in NMR.  These distributions are created by sampling from a normal distribution of the errors on each matrix element and calculating the singular values of the sampled matrix. There are 10,000 samples in each plot and the histogram bin size is $0.005$.  The cumulative of each distribution is included to guide the reader in estimating the integral of portions of the distributions. We deduce that the correlation matrix of the initial state has a rank of 1 and the correlation matrix of the final state has a rank of at least 3, which is to say that we did not detect quantum discord in the initial state, but did witness the presence of quantum discord in the final state of the DQC1 computation.  Quantum discord is witnessed in $\rho_{pps}$, which implies that the physical state $\rho_{N\!M\!R} = (1-\alpha)I / 2^N+ \alpha\rho_{pps}$ also has non-zero discord, regardless of the polarization $\alpha$.}
\label{results_f}
\end{figure}

\textbf{Results}:  We tested for the existence of quantum discord at both the beginning and final stages of the DQC1 algorithm.  To determine the rank, we choose to first measure four columns of the correlation matrix, denoting it $R_{trunc}$, the truncated correlation matrix, and test to see if the rank is greater than two. If not, we will measure additional columns until we find a rank greater than two or reach full tomography.  
To estimate the spread of the singular values as a result of uncertainties in the experiment, we perform a Monte Carlo sampling to determine which singular values can be reliably distinguished from zero.  Shown in Figure~\ref{results_f} are the distributions of singular values for the (a) initial and (b) final states. The sampling results are binned and normalized by the total number of samples and bin width to produce a distribution of the relative occurrence. The cumulative of each distribution is included to guide the reader in estimating the integration of portions of the distributions. 

In the initial state, shown in Figure \ref{results_f}a, there is one clear non-zero singular value and three others that are very close to zero.  Because the rank was not determined to be greater than 2, all columns of the correlation matrix were measured.  The histogram is generated by sampling from 1000 random combinations of four columns, always including the first column and sampling from their errors ten times each, giving a total of $10,000$ samples. 

To measure the final state we started by measuring the columns corresponding to $B_m = III, IZI, IIZ, IZZ$; recall that the four rows refer to $A_n = \{I, X, Y, Z\}$. The truncated correlation matrix is shown in equation (\ref{results_matrix}), and for example, the $(2,3)$ element corresponds to the observable $XIIZ$ and is equal to $-0.13$. Each matrix element is calculated using linear inversion of variables fitted directly from the NMR spectrum.  In order to obtain these results, four instances of the experiment were required, each with a different readout pulse and observing a different spin.  The uncertainties reported in equation (\ref{results_matrix}) are propagated through linear inversion from spectral peak-fitting, and correspond to 68.2\% confidence level in the results of the fitting process. 
\begin{eqnarray}
R_{trunc} &=& \left( \begin{array}{cccc} 1 & -0.01&  0.00& -0.01 \\ 0.10 & -0.34 & -0.13 &0.25 \\ 
0.17 & 0.38 & 0.04 & 0.26 \\ 0.01 & 0.08 & -0.01 & 0.02  \end{array} \right) \label{results_matrix}\\ 
&\pm& \left( \begin{array}{cccc} 0 & 0.007 &  0.01 & 0.007 \\ 0.05 & 0.05 & 0.05 &0.05 \\ 
0.05 & 0.05 & 0.05 & 0.05 \\ 0.04 & 0.007 & 0.007 & 0.007  \end{array} \right)  \nonumber 
\end{eqnarray}
Note that the $IIII$ term has no error associated with it as it is assumed to be exactly 1 since $\mbox{tr}(\rho) = 1$.  All values in the correlation matrix are measured as a fraction of $\alpha$ by comparison with a known reference state.  Upon inspection of the results matrix in equation (\ref{results_matrix}) it can be seen that there are likely at least three linearly independent columns (columns 1, 2, and 4) due to the relative signs of the largest elements.  This is confirmed by examining the singular values of $R_{trunc}$, the distribution of which is shown in Figure \ref{results_f}b.  Because three non-zero singular values were found immediately no additional columns of $R$ were measured. Therefore, non-zero discord is reliably detected in the final state of the DQC1 algorithm, even though simulations indicate the amount of discord present is $5.4 \times 10^{-11}$.


Conclusion: Since the polarization in the DQC1 algorithm does not affect the presence of discord in the full quantum state, we successfully detected a very small amount of quantum discord in a bulk ensemble, highly mixed, liquid state NMR implementation of the DQC1 algorithm.  There was no conclusive evidence for the presence of discord at the beginning of the computation, consistent with simulations indicating discord is generated in the DQC1 algorithm.  Whether or not these correlations are of use in the computation is a question of great interest and will shed more light on our understanding of quantum systems.


\begin{acknowledgments}
G.P.~would like to thank M.~Ditty for his technical expertise with the spectrometer.  This work was funded by NSERC, QuantumWorks, and CIFAR.  

Upon completion of this manuscript, we became aware of 
a concurrent independent study~\cite{R.-Auccaise:2011uq} that corroborates our conclusions, but uses a different method.
\end{acknowledgments}


\end{document}